\documentclass[sigconf]{acmart}
\usepackage{graphicx} 

\usepackage{tcolorbox}
\newtcolorbox{mytextbox}[1][]{%
  sharp corners,
  colback=white,
  attach title to upper,
  #1,
  fontupper=\normalsize
}

\usepackage{todonotes}

\begin{document}

\title[Vibe Coding in Practice]{Vibe Coding in Practice: Motivations, Challenges, and a Future Outlook -- a Grey Literature Review}

\author{Ahmed Fawzy}
\affiliation{%
\institution{Massey University}
\country{New Zealand}}
\email{ahmed.mohamed.6@uni.massey.ac.nz}

\author{Amjed Tahir}
\affiliation{%
 \institution{Massey University}
 \country{New Zealand}}
\email{a.tahir@massey.ac.nz}

\author{Kelly Blincoe}
\affiliation{%
  \institution{The University of Auckland}
  \country{New Zealand}}
\email{k.blincoe@auckland.ac.nz}

\renewcommand{\shortauthors}{Fawzy et al.}


\begin{abstract}
\AItools{} are transforming software development, especially for novice and non-software developers, by enabling them to write code and build applications faster and with little to no human intervention. Vibe coding is the practice where users rely on \AItools{} through intuition and trial-and-error without necessarily understanding the underlying code. Despite widespread adoption, no research has systematically investigated why users engage in vibe coding, what they experience while doing so, and how they approach quality assurance (QA) and perceive the quality of the \aigc{}. 

To this end, we conduct a systematic grey literature review of \NumberIncluded{}  practitioner sources, extracting \NumBehavioralUnits{} firsthand behavioral accounts about vibe coding practices, challenges, and limitations.
Our analysis reveals a speed–quality trade-off paradox, where vibe coders are motivated by speed and accessibility, often experiencing rapid ``instant success and flow'', yet most perceive the resulting code as fast but flawed. QA practices are frequently overlooked, with many skipping testing, relying on the models' or tools' outputs without modification, or delegating checks back to the \AItools{}. This creates a new class of vulnerable software developers, particularly those who build a product but are unable to debug it when issues arise.

We argue that vibe coding lowers barriers and accelerates prototyping, but at the cost of reliability and maintainability. These insights carry implications for tool designers 
and software development teams. Understanding how vibe coding is practiced today is crucial for guiding its responsible use and preventing a broader QA crisis in AI-assisted development.



\end{abstract}

\newcommand{\Fix}[1]{\textcolor{black}{#1}}

\newcommand{\aigc}{AI-generated code}
\newcommand{\AItools}{AI code generation tools}
\newcommand{\vb}{vibe coding}

\newcommand{\NumInitialSearch}{\Fix{154}}
\newcommand{\NumberIncluded}{\Fix{101}}
\newcommand{\NumResultsExcluded}{\Fix{53}}
\newcommand{\NumBehavioralUnits}{\Fix{518}}

\newcommand{\NumMotivation}{\Fix{140}}
\newcommand{\NumExperiences}{\Fix{132}}
\newcommand{\NumPreceptions}{\Fix{114}}
\newcommand{\NumQApractices}{\Fix{132}}


\maketitle

\section{Introduction}
\label{intro}

Recent progress in large language models (LLMs), accessible through \AItools{}, such as GitHub Copilot and ChatGPT, is rapidly transforming software development. These tools enable developers to describe functionality in natural language and receive executable code, thereby speeding up routine work and lowering the barrier to entry for individuals with limited programming experience \cite{peng2023impact, scholl2024novice, haindl2024does}. With the use of these tools, even people without any formal training are increasingly able to develop functional applications \cite{feldman2024non}. This change represents a broader shift in developer roles, which now involve orchestrating, supervising, and integrating rather than writing every line of code \cite{smith2025spectrum, naughton2025guardian}. However, while these tools are transforming how software is created, less is known about the new coding practices emerging from their everyday use.

Within the wave of rapid adoption of \AItools{}, a new practice known as \textit{vibe coding} has emerged. Coined by Karpathy in 2025 \cite{karpathy2025vibecoding}, vibe coding is a new programming approach where users employ \AItools{} to write code by describing their desired outcome (in natural language) without fully understanding the \aigc{}. For example, a recent report noted that 25\% of Y Combinator’s Winter 2025 startups had codebases written almost entirely by \AItools{}, illustrating how quickly this practice is spreading \cite{mehta2025techcrunch}. In contrast to AI-assisted programming, vibe coding prioritizes speed and experimentation over understanding.

In this paper, we define vibe coding as the practice of using \AItools{} to produce software primarily by describing goals in natural language and iteratively prompting, while relying on minimal review of the generated code. The definition is derived from Karpathy’s original introduction of the term \cite{karpathy2025vibecoding} 
and further grounded in how practitioners themselves describe the practice in grey literature (e.g., \cite{GL005}). This definition applies across user groups, including both professional developers (who may use vibe coding to prototype quickly) and non-software developers (who may attempt to build applications without coding knowledge) \cite{willison2025improvising, edwards2025vibecoding}.

Vibe coding has introduced advantages such as accessibility, speed, and creative potential. For example, non-software developers can now build working applications, and professionals can explore ideas faster than before \cite{zviel2024good, prather2023s, prather2024widening}. However, prior research and industry commentary warn of severe limitations. Educational studies show that students often adopt \aigc{} without understanding it, risking poor learning outcomes \cite{scholl2024novice, haindl2024does}. Security research has demonstrated that \aigc{} frequently contains vulnerabilities \cite{pearce2025asleep, majdinasab2024assessing}, that has also been identified across existing projects \cite{fu2025security}. Practitioners have also highlighted risks of technical debt \cite{GL005, edwards2025vibecoding}. These findings suggest that while vibe coding lowers barriers and accelerates development, it also introduces concerns regarding quality, security, and maintainability.

Despite the widespread adoption in practice, there is still no research that systematically examined vibe coding as a distinct practice. Existing studies either focus on general \AItools{} use without isolating the intuition-driven trial-and-error style that defines vibe coding \cite{vaithilingam2022expectation, peng2023impact, kruse2024can,kashif2025developers}. To address this gap, we conduct a systematic grey literature review (GLR) to analyze firsthand behavioral accounts of vibe coding documented in blogs, forums, media articles, and other publicly available sources. In this review, we attempt to answer the following four research questions: 

\noindent \textbf{RQ1: What are the motivations behind vibe coding?}\\ This question examines why users opt to vibe code, delving deeper into their motivations for choosing to code in this manner (such as speed, ease of use, creativity, learning, or accessibility). We also explain the contexts in which users find the practice valuable (such as when rapid prototyping, working on personal projects, or empowering non-software developers).
    
\noindent \textbf{RQ2: What is the user experience while vibe coding?} \\
In this question, the focus is on users’ hands-on experiences while vibe coding, exploring what worked well and what went wrong. 
 We aim to capture how users actually work with \AItools{} during vibe coding.

\noindent \textbf{RQ3: What is the perception of the \aigc{} quality with vibe coding ?}  \\
The question investigates how users perceive the outputs they produce with \AItools{}. 
It captures how users judge the usefulness and reliability of the code they obtain.
   
\noindent \textbf{RQ4: What quality assurance (QA) practices are applied when vibe coding?} \\ 
Here we explore how users check or manage the quality of \aigc{}, and whether they carefully review 
it before acceptance, or instead fully trust the output without verifying its correctness and quality.
    

By synthesizing insights from diverse practitioners and settings, our study offers a grounded account of how vibe coding is practiced in real-world contexts. We find that users are primarily motivated by speed and accessibility, but their reliance on minimal review often results in fragile or error-prone code. These insights offer implications for tool designers and software teams (e.g., designing tools that encourage review and validation rather than uncritical acceptance of \aigc{}).

\section{Related Work}

Research on \AItools{} has grown rapidly, focusing on different user groups and concerns. While these studies provide valuable insights, no direct investigation has been done on vibe coding as a distinct practice.


Some studies have examined how students and novices use \AItools{} in learning contexts. Prather et al. \cite{prather2023s, prather2024widening} found that students often accept AI suggestions without deeply considering or understanding them, leading to confusion when errors arise. Zviel-Girshin et al. \cite{zviel2024good} studied a full class of beginner software developers and reported that while students felt more confident using \AItools{}, they often did not fully grasp the concepts behind the \aigc{}. Sheard et al. \cite{sheard2024instructor} interviewed instructors, many of whom expressed concern that students might submit AI-generated work without understanding it. Zi et al.\cite{zi2025would} similarly found that CS1 students struggled to understand LLM-generated code, with only 32.5\% success in comprehension tasks due to unfamiliar coding styles, automation bias, and limited experience. Overall, these findings suggest that while \AItools{} can support confidence and accessibility, over-reliance may harm learning outcomes \cite{scholl2024novice, haindl2024does}.

Other work has examined individuals with no formal programming training. Feldman and Anderson \cite{feldman2024non} studied how non-software developers use \AItools{} to generate working code. While they were able to produce basic programs, they struggled to articulate their intent in prompts clearly and to verify whether the resulting \aigc{} was correct. These findings suggest that while \AItools{} open access to new user groups, significant barriers remain in effectively prompting and validating AI outputs.

Ferino et al. \cite{ferino2025junior} investigated how junior developers adopt GitHub Copilot, finding that many relied on trial-and-error and accepted suggestions without full understanding. At the professional level, Barke et al. \cite{barke2023grounded} found that experienced developers used \AItools{} in two main ways: (1) to accelerate code they already knew how to write, and (2) to explore unfamiliar ideas, while generally testing results carefully. Vaithilingam et al. \cite{vaithilingam2022expectation} found that developers enjoyed using Copilot but often struggled to fix generated code when bugs appeared. Peng et al. \cite{peng2023impact} reported productivity benefits, particularly for less experienced developers, though the quality of \aigc{} was not evaluated. Together, these studies suggest that developer experience plays a role in how carefully \AItools{} are used.

In prompt engineering, Kruse et al. \cite{kruse2024can} investigated how developers craft prompts for code generation and how their experience influenced outcomes. While this sheds light on the importance of prompt design, it does not specifically examine vibe coding, where trial-and-error prompting and intuition often dominate.

Several studies highlight risks associated with adopting \AItools{} without sufficient code review. Pearce et al. \cite{pearce2025asleep} found that approximately 40\% of Copilot outputs (out of 1,689 programs) contained security vulnerabilities. Majdinasab et al. \cite{majdinasab2024assessing} showed that even with additional safety layers, insecure code was still frequently produced. Fu et al. \cite{fu2025security} further identified security weaknesses in \aigc{} across GitHub projects. These findings show that while \AItools{} increase accessibility and speed, the resulting code may introduce significant risks if not carefully checked.

Beyond technical outcomes, some studies explore how developers perceive the role of \AItools{}. Kuhail et al. \cite{kuhail2024will} examined developers’ views on ChatGPT, focusing on job security, role changes, and whether AI would replace or augment programming work. 
Weisz et al. \cite{weisz2025examining} studied IBM’s use of an AI code assistant (watsonx Code Assistant) and found that, although many developers reported productivity gains, these benefits were uneven and often raised concerns about authorship, responsibility, and loss of skills. While relevant to understanding adoption, this research does not investigate how developers actually check or test \aigc{} in practice.

In summary, prior research has examined educational contexts, non-software developers, juniors, professionals, prompting, security, and developer perceptions. However, none directly addresses vibe coding as an intuition-driven trial-and-error practice distinct from general AI-assisted software development. This gap motivates our GLR, which synthesizes firsthand accounts of vibe coding behaviors across diverse user groups.

\section{Methodology}
\label{Meth}


Grey literature encompasses sources that are not published in traditional peer-reviewed venues, including blogs, technical reports, and web articles. In software engineering, practitioners often share their experiences and practices via these online channels rather than academic journals \cite{KAMEI2021106609,kamei2019use}.

Several studies have demonstrated that GL is becoming a significant source of evidence in SE. 
For example, a review found that approximately one in five secondary studies already utilize GL, as it incorporates perspectives from real practitioners that may not be evident in peer-reviewed research \cite{KAMEI2021106609}. Kamei et al.~\cite{kamei2019use} also argue that GL reviews are useful because they capture insights from materials that practitioners themselves read (like blog posts and technical reports). 

\begin{figure}[h]
    \centering
    \includegraphics[width=0.90\linewidth]{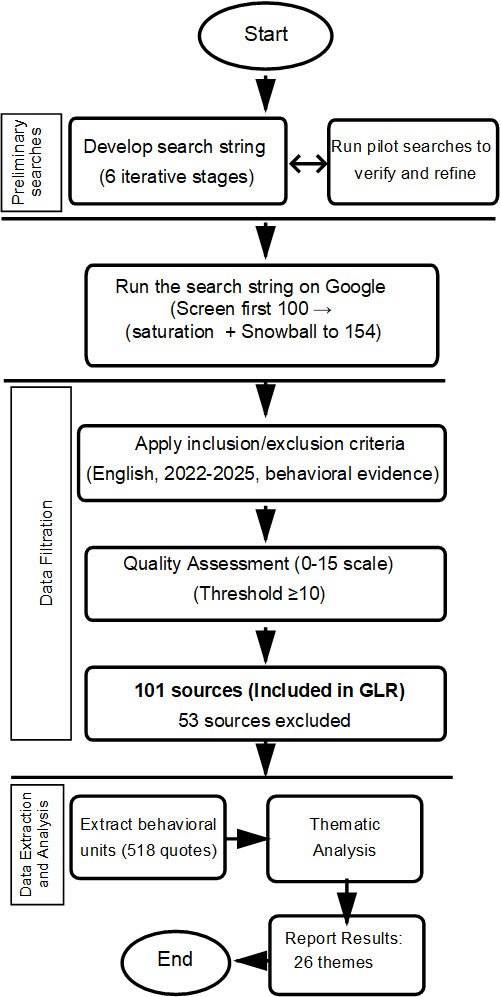}
    \caption{Grey Literature Process}
    \label{fig:GLprocess}
\end{figure}

\subsection{Search Strategy}
\label{GL-and-Search}
In conducting this review, we have closely followed the review guidelines proposed by Garousi et al.~\cite{garousi2019guidelines}.
As grey literature is spread across the web, we used \emph{Google} as our search engine. 
Prior work has shown that Google is particularly effective for discovering diverse forms of grey literature across blogs, forums, and media posts \cite{adams2016searching,adams2016synthesising,tahir2023test}. We also applied backward snowballing (following links and references inside included sources) to find further relevant sources (see Section~\ref{FinalSearchFiltration}). We discuss our search string development, data extraction, and analysis steps below.  Figure \ref{fig:GLprocess} outlines the full steps of the GL process.

\subsection{Search String}
\label{SearchStrDev}
We followed an iterative, multi-stage process to develop an effective search string. 
Our goal is to maximize both recall and practical relevance. We explain our string development steps below: 

\subsubsection{Stage 1: Problem Framing and Concept Definition}
Vibe coding refers to situations where software developers and other users (e.g., novices / non-software developers) rely on \AItools{} based on intuition and trial and error, often without a thorough understanding of the code or conducting a rigorous review and testing. This distinguishes it from more structured or expert-guided AI-assisted programming. Therefore, our search needed to focus on behavioral indicators and practitioner phrasing, not just technical descriptions.

\subsubsection{Stage 2: Initial Candidate Terms}
We began with domain-informed candidate terms commonly seen in early articles, consisting of the following terms:
\begin{mytextbox}
(``AI-assisted programming'' OR ``AI code generation'' OR ``prompt-based programming'' OR ``LLM-assisted development'') $+$  tool names (e.g., ``ChatGPT'' or ``Copilot'') $+$ role (e.g., ``developer'' OR ``student'' OR ``non-programmer'')
\end{mytextbox}

However, this term combination had limited success. The search retrieved mostly unrelated sources (e.g., more technical AI articles, promotional content, or conceptual content) and missed the primary target of our GLR. The first 100 results (a manageable representation of relevant grey literature \cite{briscoe2024alternative,adams2016searching}) of this stage were screened against our inclusion and exclusion criteria (see Section~\ref{IncExc}), but since they did not contain firsthand evidence of vibe coding behaviors, they were excluded (screened 100, ~8\% relevant). This stage nevertheless helped us refine our search strategy by clarifying which terminologies were too broad or non-practitioner-focused.

\subsubsection{Stage 3: Pivot to Practitioner Language}
We then shifted focus to how practitioners describe AI coding tools in everyday usage. We noticed that several terms are frequently used, including:
``coding with AI'', ``AI writes code'', ``prompting for code'', ``using AI to code'', ``programming with an AI assistant''. We also incorporated the term \texttt{``vibe coding''} directly, as it had become widely adopted following its introduction in 2025~\cite{karpathy2025vibecoding}. In this stage, we screened 100 sources, with approximately 50\% considered relevant.

\subsubsection{Stage 4: Iterative Expansion and Saturation}
As part of an iterative design and test process, we expanded the candidate string to capture semantically adjacent behaviors. We added terms such as \texttt{``LLM-assisted coding''}, \texttt{``AI programming tool''}, and \texttt{``AI pair programmer''}, which returned relatively few additional relevant hits  (~12\%) when we screened 100 results, based on their appearance in pilot searches and known grey literature sources (Quasi-Gold Standard see Section \ref{QGold}). Each term was tested for recall and specificity using Google search, and retained only if it increased the relevance of retrieved documents.

\subsubsection{Stage 5: Final Search String}
\label{FinalSearchFiltration}
The final version of the search string retrieved all items in the quasi-gold standard set and produced strong recall without adding excessive unnecessary information. We initially screened the first 100 results of the final search string, then continued until we reached thematic saturation (additional screening no longer surfaced substantially new related sources). Backward reference (Snowballing) chaining was applied to the first 100 results of the final search string retrieved sources, following the guidelines of Wohlin et al \cite{wohlin2012experimentation}. This involved tracing hyperlinks, named references, or cited resources within blogs and forums (e.g., \cite{GL005,GL013},\cite{GL034,GL055}). This leads to a total of \NumInitialSearch{} sources.
\textbf{The  String:}
 \begin{mytextbox}
\normalsize
``vibe coding'' OR ``coding with AI'' OR ``AI writes code'' OR ``AI code assistant'' OR ``LLM-assisted coding'' OR ``AI programming tool'' OR ``prompting for code'' OR ``using AI to code'' OR ``AI pair programmer'' OR ``programming with an AI assistant''
 \end{mytextbox}


\subsubsection{Stage 6: Quasi-Gold Standard Validation and Pilot Search}
\label{QGold}
To ensure the effectiveness of the final string, we employed a \textit{quasi-gold standard} evaluation approach. Additionally, this involved testing each search iteration against a curated set of known relevant grey literature sources, posts we had previously identified as clear examples of vibe coding behavior. These sources were identified during our pilot search stages and met all inclusion criteria (see Section~\ref{IncExc}) with high quality scores (see Section~\ref{QAasst}). The quasi-gold set included firsthand accounts involving AI prompting, minimal oversight, trial-and-error acceptance, or skipped QA practices. Specifically, it comprised: Andrej Karpathy, who coined the term \emph{vibe coding} in early 2025 and demonstrated it by developing a restaurant menu application using this approach \cite{GL039}, including clear experiential
reflections on the process and deployment; Simon Willison, who emphasized that not all
AI-assisted programming is vibe coding \cite{GL005}, and described his own prototyping
workflow and personal experiences using LLMs to vibe code \cite{GL013}; and Maxime Najim,
An experienced software engineer, whose reflective post documented the intentional and risky use of vibe coding in practice \cite{GL088}; and Tyler Shields, an early practitioner account that described vibe coding as relying entirely on LLMs without verification \cite{GL009}. This validation process increased our confidence that the final search string would capture
the full range of behavioral evidence needed to address our research questions, and was
aligned with the guidelines of Garousi et al.~\cite{garousi2019guidelines}.

\subsection{Inclusion and Exclusion Criteria}
\label{IncExc}
We applied a set of inclusion and exclusion criteria. Our inclusion criteria are:
\begin{itemize}
    \item English-language sources
    \item Sources published between 2022 and 2025
    \item Relevance to the research questions
    \item Identifiable author or publishing entity
    \item Publicly accessible full text
    \item Behavioral evidence of vibe coding, including: AI-generated code used as-is or with light edits, prompting and reprompting, minimal or no testing, review, or improvement
    \item Critical or negative reflections are accepted if based on firsthand experience with vibe coding
\end{itemize}

We excluded material that met any of the following exclusion criteria:

\begin{itemize}
    \item Duplicates or mirrored content
    \item Irrelevant topics or purely technical focus
    \item Promotional content
    \item Undated or anonymous sources
    \item Tool descriptions without user insight (e.g., Copilot/ChatGPT feature summaries)
    \item Very short content (e.g., tweets or headlines without commentary)
    \item Abstract speculation or general opinions unsupported by observable vibe coding actions such as prompting, accepting, skipping QA.
    \item No firsthand behavior involving vibe coding, such as:
    \begin{itemize}
        \item No prompting or AI code generation
        \item No code acceptance or editing
        \item Experienced developers applying critical oversight without skipping QA
    \end{itemize}
\end{itemize}

\subsection{Quality Assessment}
\label{QAasst}
The quality assessments were mainly conducted by the first author, with each source scored from 0 to 3 across five quality dimensions, following Garousi et al.'s GL guidelines \cite{garousi2019guidelines} with a maximum score of 15. These five quality criteria (Authority, Evidence, Objectivity, Currency, and Purpose) were adapted from Garousi et al.'s \cite{garousi2019guidelines} framework for grey literature quality assessment, which has been widely adopted in software engineering GLR \cite{soldani2018pains, KAMEI2021106609}. To increase reliability, a subset of sources was cross-validated by another co-author. Sources scoring $\geq 10$ were retained for further analysis. The full scoring descriptions, including the results and justification for each article quality score, have been included in our dataset \cite{fawzy_2025_17188020}.

\noindent \textbf{Scoring Criteria:}
\begin{itemize}
    \item \textbf{Authority:} Author identity and credibility (e.g., known developer, reputable organization)
    \item \textbf{Evidence:} Use of examples, rationale, or empirical support
    \item \textbf{Objectivity:} Neutral and balanced presentation of ideas
    \item \textbf{Currency:} Published between 2022 and 2025
    \item \textbf{Purpose:} Intended to inform, reflect, or explain (not promote or market)
\end{itemize}

\subsection{Data Filtration:}
\label{DataFilterExtract}
Our search string retrieved \NumInitialSearch{} grey literature sources (e.g., blog posts, online articles, and technical reports), of which \NumberIncluded{} met our inclusion criteria and quality thresholds. All other \NumResultsExcluded{} sources were excluded. Among the exclusions, 40 sources were excluded because they did not reach the minimum quality threshold (QA score $\geq 10/15$), and 13 were excluded based on the inclusion and exclusion criteria (see Section~\ref{IncExc}). For example, \cite{GL003} (a Wikipedia entry), \cite{GL017} (a promotional Replit blog post), and \cite{GL016} (a satirical news commentary in The Register) were excluded for not meeting our quality thresholds for behavioral detail/provenance. By contrast, \cite{GL010} (an explanatory article in The Conversation), \cite{GL011} (a Cloudflare vendor learning page), and \cite{GL012} (a Google Cloud conceptual overview) were likewise excluded under our inclusion/exclusion criteria and for lacking firsthand behavioral evidence (no prompting, code acceptance, or QA evidence).



\subsection{Behavioral Unit Extraction}
\label{BehavioralUnits}

To move from raw articles to analyzable data, we systematically extracted
\NumBehavioralUnits{} \textit{behavioral units} from the \NumberIncluded{} included sources. A behavioral unit is a single coded instance that captures something relevant to the RQs. This may be a practitioner describing their own vibe coding, or an author/reporter documenting or commenting on vibe coding practices. 
Extraction was performed at the quote level rather than the article level, coding each instance of behavior (motivation, experience, perceptions of code quality, or QA practice) separately. This ensured that multiple distinct behaviors within the same article (e.g., a motivation to save time and a later description of skipping QA) were each represented as separate units in our dataset.


Each behavioral unit was extracted, classified and recorded in an external spreadsheet with the following fields:
\begin{itemize}
    \item \textbf{Verbatim Quote:} the exact wording from the source.
    \item \textbf{Attribution:} who expressed the behavior (author or quoted user)
    \item \textbf{RQ Mapping:} whether the unit described motivation (RQ1), experience (RQ2), perceptions of code quality (RQ3), or QA practice.
    \item \textbf{Interpretation:} a short analytic summary of what the behavior revealed
    \item \textbf{Metadata:} user type, tool mentioned, vibe coding definition (if given), and notes
\end{itemize}
Behavioral unit extraction was conducted mainly by the first author. To increase reliability, a subset of units was cross-validated by another co-author. This process yielded \NumMotivation{} motivation units (RQ1), \NumExperiences{} experience units (RQ2), \NumPreceptions{} perception of code quality units (RQ3) and \NumQApractices{} QA practice units (RQ4). These units formed the raw data corpus for our subsequent thematic analysis (Section~\ref{DataAnalysis}).

\subsection{Data Analysis}
\label{DataAnalysis}
We used thematic analysis to analyze the extracted behavioral units of grey literature data, following the procedures recommended by Braun \& Clarke \cite{braun2006using}. 
Behavioral insights were organized and coded, guided by our research questions, according to patterns in motivations, experiences, perceptions of code quality, and quality assurance practices related to vibe coding. The data analysis steps, including theme development, were mainly conducted by the first author. To increase reliability, reviews and discussions with other co-authors were carried out. We followed a set of steps to develop the themes, explained below:
\begin{itemize}
   \item \textbf{Familiarization with the Data:}
We read through all the extracted \NumBehavioralUnits{} behavioral units multiple times to understand patterns. For example, \textit{``I just trust it works.''} \cite{GL023} and
\textit{``They’re accepting the responses in a very trusting way.''}\cite{GL035} suggests a common pattern of uncritical trust in AI-generated code without any checks.
  \item \textbf{Generating Initial Codes:}
 We then labeled each behavioral unit with a potential theme. For example, \textit{``Copy and paste them in…usually, that fixes it.''} we labeled it as \textit{``Reprompting Instead of Debugging''}. We set the potential theme ``Speed'' to \textit{``Alex Finn famously created a Call of Duty style shooter game in just 87 minutes using AI tools.''} \cite{GL050}. These initial codes summarize the behaviors, but they do not group them yet.
  \item \textbf{Searching for Themes:}
 We grouped similar initial codes (potential themes) into larger candidate themes. This means we started clustering behaviors that reflected the same idea or pattern. The potential themes such as ``Uncritical Trust'' \cite{GL094}, ``False Confidence'' \cite{GL080}, and ``Uncritical Security Trust'' \cite{Gl119} were grouped into the final theme ``Uncritical Trust''.
  \item \textbf{Reviewing Themes:}
 We examined whether our potential themes were consistent throughout the entire dataset. Sometimes we split, rename, or reassign behaviors to better-fitting themes. For example, some quotes originally coded as ``Skipped QA'' were actually different in behavior. For instance, \textit{``Businesses often trust AI-generated code uncritically, leading to vulnerabilities and technical debt'}' \cite{GL050} were reassigned to the final theme ``Uncritical Trust'' as they choose to trust without understanding.
  \item \textbf{Defining and Naming The Final Themes:}
For each theme we identified, we wrote a clear description that explained what the theme represented and how it was different from the other themes. A consolidated overview of all theme definitions is provided in Table~\ref{tab:theme-definitions}.
\item \textbf{Producing the Report:}
Finally, we prepared the results for presentation. Each final theme is reported in Section~\ref{results} together with: (its definition, frequency statistics, and illustrative explanations).
\end{itemize}

Table~\ref{tab:theme-definitions} summarizes the final theme definitions used to code behavioral units across RQ1 (motivations), RQ2 (experiences), RQ3 (perceptions of \aigc{} quality) and RQ4 (quality assurance practices). Figure \ref{fig:Allthemes} displays all the vibe coding behavior themes we have constructed, along with their corresponding frequencies.

\begin{table*}[]
\caption{Description of the identified vibe coding themes}
\label{tab:theme-definitions}
 \resizebox{\linewidth}{!}{%
\begin{tabular}{ll}
\hline
\textbf{Theme} & \textbf{Description} \\ \hline
\multicolumn{2}{l}{\textit{Motivation for vibe coding (RQ1)}} \\ \hline
Speed \& Efficiency & The ability to produce working software much \textbf{ faster, reducing development cycles} from weeks to only hours. \\
Accessibility \& Empowerment & Enabling \textbf{Non-Software Developers} ability to create applications by describing goals in natural language. \\
Learning \& Experimentation & Using LLM coding models as a \textbf{tutor or sandbox} to explore new coding tools, concepts, and frameworks. \\
Creative Exploration & Turning imaginative or artistic \textbf{ideas into functional projects} with AI support.\\
Fast Prototyping & Quickly assembling MVPs or \textbf{demos to test feasibility}, concepts, or market interest.\\
Reducing Mental Effort & Offloading syntax and boilerplate to AI, \textbf{easing the developer’s cognitive load}. \\
Frustration Avoidance & \textbf{Bypassing programming pain points} e.g., repetitive debugging, setup issues etc. \\
Escape Complexity & Avoiding difficult design or architecture work by \textbf{delegating detail to AI}. \\
Curiosity or Play & Engaging casually with \textbf{AI for fun}, experimentation, or open-ended exploration. \\ \hline
\multicolumn{2}{l}{\textit{Experience with vibe coding (RQ2)}} \\ \hline
Instant Success \& Flow & Experiencing \textbf{rapid, seamless progress} that creates momentum, leading to satisfaction.\\
Prompt Struggle \& Iteration & \textbf{Refining prompts through repeated trial-and-error} until usable results appear. \\
Code Breakdown or Abandonment & \textbf{Abandoning projects} when AI outputs become too buggy or complex. \\
Fun \& Creative Satisfaction & \textbf{Enjoying the excitement of fast software development}, though the novelty can fade. \\
AI Hallucinations & Facing \textbf{false, inaccurate, or misleading code suggestions} that look plausible but fail in execution or introduce bugs. \\
Confusion or Misunderstanding & \textbf{Misaligned prompts and outputs} causing frustration and loss of trust \\ \hline
\multicolumn{2}{l}{\textit{Code Quality Perception (RQ3)}} \\ \hline
Fast but Flawed & Useful for \textbf{quick tasks but unsuitable for production} deployment. \\
Fragile or Error-Prone & Containing \textbf{ hidden bugs}, inconsistencies, or potential security risks. \\
Sloppy or Low Maintainability & Functional but \textbf{poorly structured}, undocumented, and hard to extend. \\
Prototype-Ready Only & Adequate for \textbf{demos and proofs-of-concept} but not for long-term systems. \\
High Quality \& Clean & Actually \textbf{well-structured}, and close to production quality \\
Misleading Confidence & Appearing reliable while concealing deeper flaws, \textbf{creating false trust}. \\ \hline
\multicolumn{2}{l}{\textit{QA Practices (RQ4)}} \\ \hline
Skipped QA & \textbf{Bypassing structured testing} or review, relying only on execution success. \\
Manual Testing or Edits & Applying systematic checks (\textbf{manual code reviews}) and edits before adopting AI outputs. \\
Uncritical Trust & Accepting \aigc{} without validation or verification, \textbf{based on the assumption that it works}. 
\\
Delegated QA to AI & Relying on the \textbf{\AItools{} to detect} and correct \textbf{its own mistakes}. \\
Reprompting Instead of Debugging & \textbf{Feeding errors back into \AItools{}} rather than fixing them manually. \\
Run-and-See Validation & \textbf{Running code} to check if it works, equating success with correctness. \\
QA Breakdown or Confusion & \textbf{Failing to validate} the outputs due to complexity or lack of clarity. \\ \hline
\end{tabular}%
 }
\end{table*}

\section{Results}
\label{results}

\begin{figure*}[]
    \centering
     \resizebox{0.70\linewidth}{!}{%
\includegraphics[width=\linewidth]{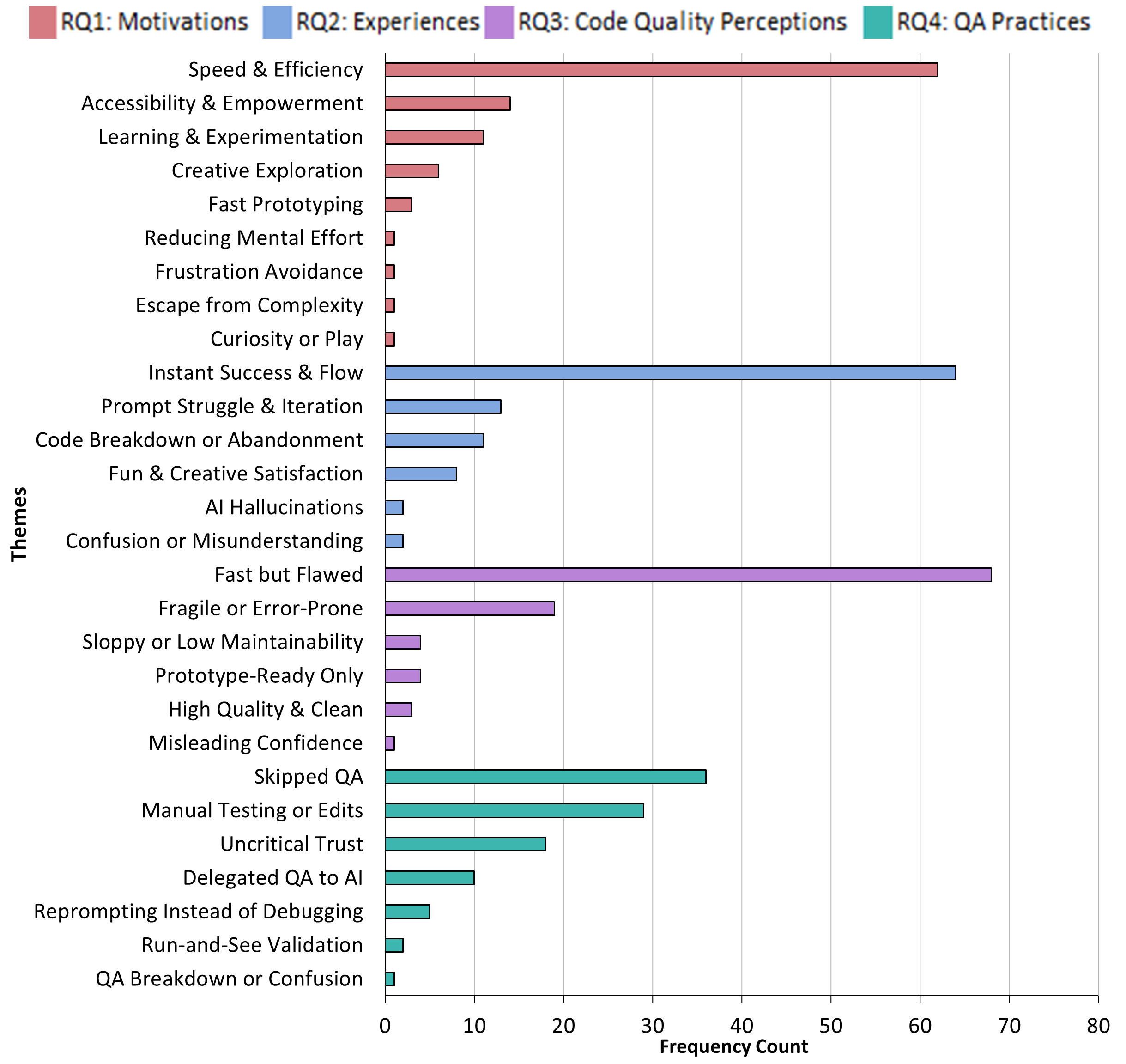}}
    \caption{The identified themes from the collected sources}
    \label{fig:Allthemes}
\end{figure*}

From these \NumberIncluded{} sources, we identified \NumBehavioralUnits{} behavioral units, which we coded into the themes defined earlier (see Table~\ref{tab:theme-definitions} and Figure~\ref{fig:Allthemes}), including motivations (\NumMotivation{} units, RQ1), experiences (\NumExperiences{} units, RQ2), perceptions of code quality (\NumPreceptions{} units, RQ3), and QA practices (\NumQApractices{} units, RQ4).

For each RQ, we identified several distinct themes, which are summarized with definitions in Table~\ref{tab:theme-definitions}. Due to space limitations, we describe in detail only themes that account for at least 10\% of the behavioral units for that RQ. Less frequent themes are still included in Table~\ref{tab:theme-definitions} and distribution tables (Tables~\ref{tab:motivations}–\ref{tab:qa}).


\subsection{Motivations for Vibe Coding (RQ1)}

There were \NumMotivation{} behavioral units related to motivation, from which we identified nine distinct themes through thematic analysis. 
A summary of those motivational themes is shown in Table~\ref{tab:motivations}.
A distribution of the motivational themes, including their frequencies and percentages, is shown in Table~\ref{tab:motivations}.

\begin{table}[htbp]
\centering
\caption{Distribution of Motivations for Vibe Coding (n=140)}
\label{tab:motivations}
\begin{tabular}{lcc}
\toprule
Theme & Frequency & Percentage \\
\midrule
Speed \& Efficiency & 87 & 62\% \\
Accessibility \& Empowerment & 20 & 14\% \\
Learning \& Experimentation & 15 & 11\% \\
Creative Exploration & 8 & 6\% \\
Fast Prototyping & 4 & 3\% \\
Reducing Mental Effort & 2 & 1\% \\
Frustration Avoidance & 2 & 1\% \\
Escape Complexity & 1 & 1\% \\
Curiosity or Play & 1 & 1\% \\
\bottomrule
\end{tabular}
\end{table}

\noindent \textbf{Speed \& Efficiency}: We found that the most common motivation theme for vibe coding (62\%) is speed \& efficiency, with vibe coders highlighting rapid development. Practitioners consistently described how \AItools{} enabled them to produce working software in dramatically less time, often hours instead of weeks (e.g., \cite{GL002,GL005}). For example, one source reported that they have built over 140k lines of code (LOC) with tests and documentation in just under 15 days using vibe coding practices \cite{GL051}. This acceleration was valued not only by beginners but also by experienced developers and even organizations \cite{GL006}. Across these groups, speed was not just about coding faster; it translated into tangible productivity outcomes. For instance, one source expects that the company will reduce manual testing time by 25\%, achieve complete test coverage on projects 20\% faster, and fix significantly more bugs early in the development cycle \cite{GL074}.

\noindent \textbf{Accessibility \& Empowerment}: Another important motivation theme (14\%) was accessibility and empowerment, where vibe coding lowers the barrier to entry for software development. Non-software developers described being able to turn their ideas into functional applications simply by expressing them in natural language to \AItools{} (e.g., \cite{GL002,GL015}). This democratization represents a major shift in who can participate in software creation: tasks that once required teams of technical specialists can now be achieved by individuals without programming training. Evidence of this appeared not only in personal projects, but also in organizational contexts, where analysts and policy staff reported building automations or public-facing forms without relying on IT departments (e.g., \cite{GL096,GL141}). Overall, this theme highlights how vibe coding is not only about speed but also about expanding access to software creation for entirely new groups of users.


\noindent \textbf{Learning \& Experimentation}: A further motivation theme (11\%) was learning and experimentation, where vibe coding was used as both a hands-on tutor and a sandbox for trying out ideas. Practitioners described how experimenting with prompts and outputs helped them quickly build intuition about what works and what does not \cite{GL013}. Vibe coding also enabled users to explore new programming languages and frameworks without formal training, with novices and career changers reporting that it accelerated their learning and confidence in coding tasks \cite{GL039,GL049}. Even experienced developers reported using \AItools{} to familiarize themselves with unfamiliar domains or frameworks more efficiently, positioning vibe coding as a complementary way to learn through practice \cite{GL101,GL013}.

\subsection{Experiences During Vibe Coding (RQ2)}
There were \NumExperiences{} behavioral units related to experiences, from which we identified six distinct themes through thematic analysis. A summary of these themes is provided in Table~\ref{tab:experiences}. A distribution of the experience themes, including their frequencies and percentages, is shown in Table~\ref{tab:experiences}.


%

\begin{table}[htbp]
\centering
\caption{Distribution of Experiences During Vibe Coding (n=132)}
\label{tab:experiences}
\begin{tabular}{lcc}
\toprule
Theme & Frequency & Percentage \\
\midrule
Instant Success \& Flow & 85 & 64\% \\
Prompt Struggle \& Iteration & 17 & 13\% \\
Code Breakdown or Abandonment & 14 & 11\% \\
Fun \& Creative Satisfaction & 11 & 8\% \\
AI Hallucinations & 3 & 2\% \\
Confusion or Misunderstanding & 2 & 2\% \\
\bottomrule
\end{tabular}
\end{table}


\noindent \textbf{Instant Success \& Flow}: The most common experience theme (64\%) was instant success and flow, where vibe coders described the process as fast, easy, and often ``Magical.'' Practitioners reported being able to build working applications in minutes through simple conversations with \AItools{} \cite{GL045,GL032,GL121}. This immediacy created a sense of addictive momentum, with some describing it as a ``dopamine hit'' when prototypes, QA checks, and deployments came together quickly \cite{GL129}. Accounts ranged from developers building complete applications within hours \cite{GL031} to non-software developers expressing amazement at creating functional tools and even cultural translation projects without prior coding expertise \cite{GL042}.


\noindent \textbf{Prompt Struggle \& Iteration}: Not all experiences were positive. A common experience (13\%) was prompt struggle and iteration, where vibe coders needed to refine their instructions repeatedly to achieve a satisfactory result. Practitioners described going through cycles of prompt adjustment and code refinement, with some projects requiring dozens or even hundreds of iterations before the output was usable \cite{GL001,GL055}. This process highlighted the emergence of prompt engineering skills, where users learned to craft, adapt, and even collect effective prompts as reusable patterns to improve their results.


\noindent \textbf{Code Breakdown or Abandonment}: Another experience theme (11\%) was code breakdown or abandonment, where vibe coding sessions ended in failure. When \aigc{} produced outputs that were too complex, buggy, or inconsistent to fix, some practitioners reported giving up on projects altogether rather than attempting to debug the code \cite{GL015}. These breakdowns often occurred when task complexity exceeded the AI’s ability to generate reliable solutions, leading to frustration and project abandonment \cite{GL042}.

\subsection{Perceptions of the Generated Code's Quality (RQ3)}

There were \NumPreceptions{} behavioral units related to perceptions of code quality, from which we identified six distinct themes through thematic analysis.
A distribution of these code quality perception themes, including their frequencies and percentages, is shown in Table~\ref{tab:quality}.

\begin{table}[htbp]
\centering
\caption{Distribution of Perceived Code Quality (n=114)}
\label{tab:quality}
\begin{tabular}{lcc}
\toprule
Theme & Frequency & Percentage \\
\midrule
Fast but Flawed & 78 & 68\% \\
Fragile or Error-Prone & 22 & 19\% \\
Sloppy or Low Maintainability & 5 & 4\% \\
Prototype-Ready Only & 5 & 4\% \\
High Quality \& Clean & 3 & 3\% \\
Misleading Confidence & 1 & 1\% \\
\bottomrule
\end{tabular}
\end{table}


\noindent \textbf{Fast but Flawed}: We found that the most common perception of the \aigc{} quality theme is ``fast but flawed'' (68\%), where practitioners acknowledged clear trade-offs between speed and long-term quality. Vibe coders noted that while \AItools{} could quickly generate most of a solution, the remaining critical work to make code production-ready often became a challenge \cite{GL045}. They accepted these flaws as an inevitable cost of rapid development, keeping the code as long as it worked but recognizing that this created technical debt over time \cite{GL032}.


\noindent \textbf{Fragile or Error-Prone}: Another common perception (19\%) was that \aigc{} is fragile or error-prone, raising concerns about hidden issues. Practitioners cautioned that such code was often excluded from reviews or security checks, creating the risk of undetected vulnerabilities \cite{GL001}. Others emphasized that while the outputs might appear clean and functional, they could conceal subtle logic errors, performance bottlenecks, or serious security flaws that only become apparent later \cite{GL066}.

\subsection{QA Practices in Vibe Coding (RQ4)}
There were \NumQApractices{} behavioral units related to QA practices, from which we identified seven distinct themes through thematic analysis.
A distribution of these QA practice themes, including their frequencies and percentages, is shown in Table~\ref{tab:qa}.

\begin{table}[htbp]
\centering
\caption{Distribution of QA Practices in Vibe Coding (n=132)}
\label{tab:qa}
\begin{tabular}{lcc}
\toprule
Theme & Frequency & Percentage \\
\midrule
Skipped QA & 48 & 36\% \\
Manual Testing or Edits & 38 & 29\% \\
Uncritical Trust & 24 & 18\% \\
Delegated QA to AI & 13 & 10\% \\
Reprompting Instead of Debugging & 6 & 5\% \\
Run-and-See Validation & 2 & 2\% \\
QA Breakdown or Confusion & 1 & 1\% \\
\bottomrule
\end{tabular}
\end{table}


\noindent \textbf{Skipped QA}: The most common QA practice (36\%) was skipped QA, where vibe coders accepted \aigc{} without validation. Practitioners reported bypassing traditional testing entirely, for example, they did not write unit or integration tests, perform structured reviews, or systematically verify correctness beyond simply running the code.  
instead relying on whether the code executed without errors as a proxy for quality \cite{GL001}. Even experienced developers described pasting error messages back into \AItools{} and letting it generate fixes, rather than debugging or testing the code themselves \cite{GL002}.


\noindent \textbf{Manual Testing or Edits}: The second most common QA practice (29\%) was manual testing or edits, where practitioners applied careful quality control to \aigc{}. Some emphasized the risks of pushing generated code directly to production, warning that without review it could introduce bugs, security issues, or performance problems \cite{GL006}. Experienced developers described establishing stricter review protocols, treating every generated change as something that needed to be understood and verified, often supported by testing and automated checks \cite{GL088}.


\noindent \textbf{Uncritical Trust}: Another QA practice (18\%) was uncritical trust, where vibe coders believed the code worked even without checking it. As one source noted, humans tend to place more faith in generated code than is warranted, not scrutinizing it as carefully as code written by a fellow developer \cite{GL094}. This trust also extended to complex systems, with reports that people were no longer checking outputs line by line but simply accepting the responses in a very trusting way \cite{GL031}.


\noindent \textbf{Delegated QA to AI}: Another QA practice (10\%) was delegating quality checks back to the AI itself. As one source noted, users were overly reliant on the same LLMs that had introduced errors, which were also used to fix them, giving a false sense of security \cite{GL059}.

\section{Discussion}


\subsection{Key Observations}

\textbf{The Speed–Quality Trade-off Paradox:}
Our results highlight a paradox: vibe coders knowingly accept flawed \aigc{} in exchange for rapid progress. Development speed motivates all groups, but the trade-off manifests differently depending on background and experience. Non- and novice software developers often acknowledge that they can build applications quickly while conceding that \textit{``it’s not really coding''} \cite{GL002}, reflecting both excitement and awareness of possible limitations. Experienced software developers also value speed, but balance it with caution, as 29\% reported that they usually make some manual adjustments or add tests to the generated code, showing risk-aware behavior. Developers sometimes apply extensive modifications to the generated code to the level that they no longer consider it AI-generated \cite{kashif2025developers}. This divide shows that while vibe coders are willing to tolerate imperfect code for speed, only experienced users have the skills to fix problems when they arise. This observation is also reflected in the \textit{Stack Overflow Developer Survey 2025}, which reports both high adoption of \AItools{} (84\% use or plan to use them) and low trust in \aigc{} ($\sim$46\% report distrust) \cite{stackoverflow2025survey}. The implication is a divide between two groups of vibe coders: empowered novices who may remain dependent on \AItools{}, and professionals who integrate selective QA practices. 




\textit{\underline{Recommendation for Practitioners:} Use vibe coding to explore and prototype quickly, but never promote to production without adding guardrails: tests, code review, and traceable decision records (why the AI change was accepted, which checks passed, any accepted risks, and a short prompt/response ID log).}\\


\begin{figure}[]
    \centering
    \includegraphics[width=1\linewidth]{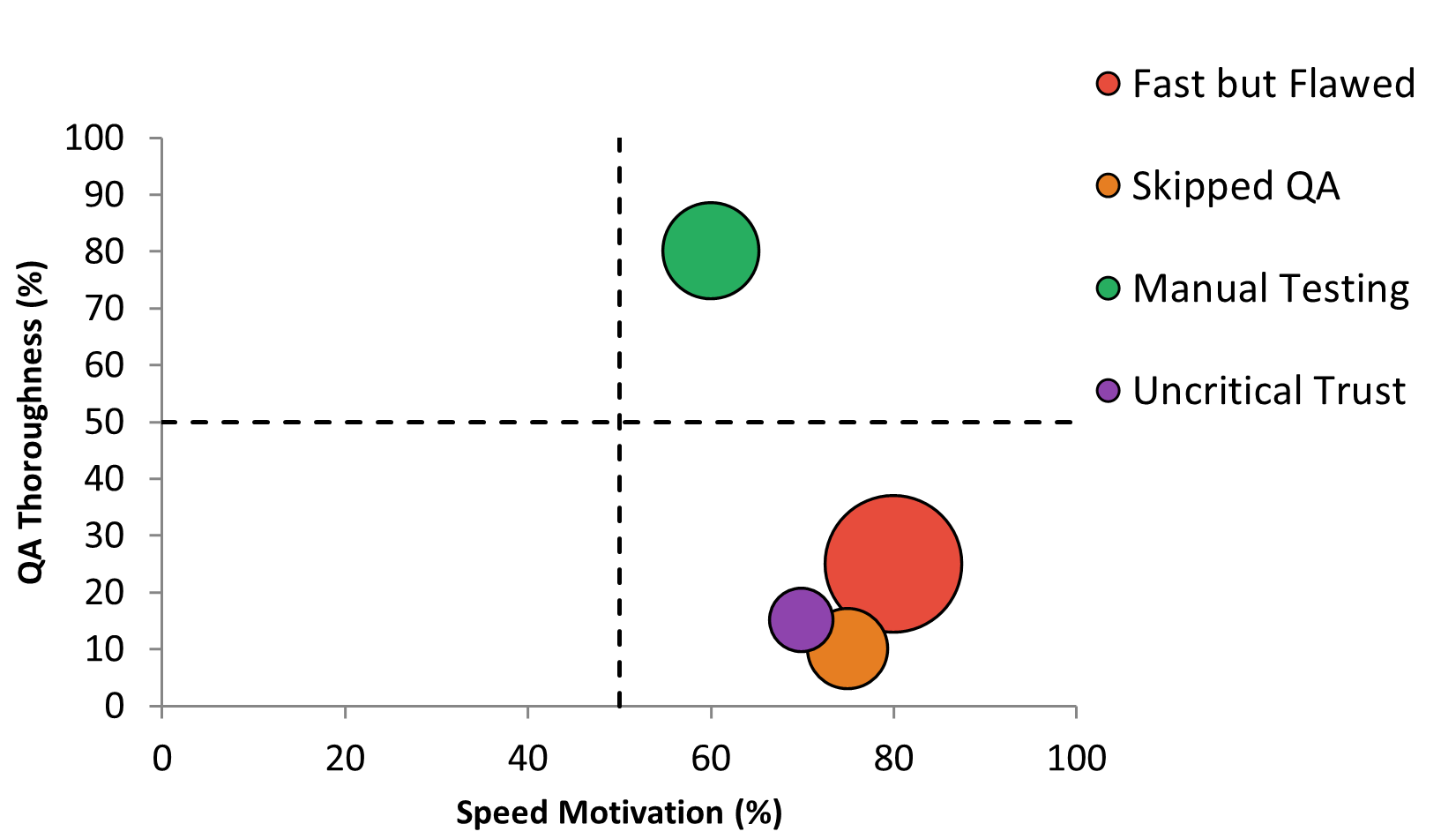}
    \caption{Speed vs QA Trade-off in Vibe Coding }
    \label{fig:SpeedQA}
\end{figure}

\noindent  \textbf{The QA Crisis in AI-Assisted Development:}
The most observed concerning issue with vibe coding is the systematic breakdown of traditional QA practices. 
A majority of QA practices reflect a departure from code verification, with practitioners commonly skipping tests, placing uncritical trust in outputs, or delegating responsibility back to the \AItools{}.
These QA concerns are possibly due to multiple factors:
\textit{Technical barriers;} e.g., the \aigc{} is cited to be difficult to debug as it can lack architectural structure \cite{GL001} 
and the contextual details software developers normally rely on, such as comments, assumptions, or information about how the code integrates with the larger system. 
\textit{Confusion:} vibe coders report confusion when attempting to understand \aigc{}.
\textit{False confidence:} the ``instant success'' experience creates illusions of correctness.
If current practices continue, vibe-coded apps may look fine but hide serious flaws. Practitioners already warn of this risk, with 19\% admitting their code is ``fragile or error-prone''. This matters because once teams get used to shipping fragile or error-prone code without proper QA, they may lower the quality bar across the whole organization. Over time, this creates a culture where untested code is considered acceptable, raising the risk of costly failures, outages, and security breaches.


\textit{\underline{Recommendation for \AItools{} designers:}} \AItools{} should incorporate lightweight verification processes and continuously remind users of the QA aspects (especially those with no formal software development experience) of the generated code. Including code (static and dynamic) analysis checks that can verify the generated code and remind developers of potential risks can be a useful tool for coders. This can be offered as a real-time indicator (e.g., performance, security concerns, missing tests) of the generated code. Tools should address the \textit{``Uncritical Trust''} issue by including features such as step-by-step code explanations (``walkthroughs''), visual diagrams, or inline explanations 
to clarify what the tool is doing and why.\\



\noindent  \textbf{New Class of Vulnerable Developers:}
One of the most notable implications of our findings is the emergence of a new class of vulnerable developers. About 14\% of vibe coders are motivated by accessibility and empowerment, with non-software developers describing how \AItools{} allow them to create applications without prior coding skills \cite{GL042}. Yet, this democratization often leaves them unprepared when problems arise. Several sources illustrate how non-software developers quickly reach dead ends when faced with bugs or technical errors that they cannot diagnose or resolve \cite{GL062, GL110}. Others highlight how uncritical trust in \AItools{} suggestions can lead to copy-paste development practices where fixes are applied without any real comprehension of their impact \cite{GL062}. In practice, this over-reliance can introduce serious risks: practitioners have documented cases of insecure systems built through vibe coding, including applications that lacked authentication, authorization, or contained hardcoded secrets \cite{GL033}. Beyond individual projects, such practices contribute to the rise of ``shadow IT'', where employees outside formal development teams build software without oversight or governance \cite{GL143}. As some experts caution, the danger is not when \aigc{} fails outright, but when it appears to work while embedding subtle vulnerabilities and technical debt \cite{GL080}. Taken together, these accounts suggest that while vibe coding lowers the barrier to entry, it also transfers significant responsibility to users who may not yet possess the necessary skills to manage it effectively.


\textit{\underline{Recommendation for Organizations and Practitioners:} Match tasks to skill and provide scaffolds (guided debugging, safe templates, and escalation paths), so newcomers learn to diagnose issues rather than outsource all QA to the AI.}
This is particularly important to avoid the novice developer trap, where failures can lock beginners into reprompt–paste loops, accepting fragile behavior when they cannot restore alignment between intent (their goal) and implementation (what the \aigc{} actually does).

\subsection{Future Work and Open Research Questions}


Further research is needed on how vibe coders’ practices shift with experience from non-software developers to novices to professionals, so that \AItools{} can adapt feedback, explanations, and safeguards to the user’s expertise.
Further empirical evidence is needed to understand if these new practices will lead to new defect and vulnerability patterns unique compared to both conventional AI-assisted (used by experienced developers) and human coding \cite{cotroneo2025human,licorish2025comparing}, to inform automated quality signals and assurance features in next-generation \AItools{}.

It is also unclear which code review practices (e.g., run-and-see checks, automated tests, AI-assisted reviews) actually work under vibe coding conditions. Research in this area can inform the design of practical, built-in QA workflows that keep pace with rapid prototyping.
Understanding how code review and QA strategies should differ for non-software developers, novices, and professionals can enable tools to enhance the quality of the generated code, providing guardrails for newcomers while supporting advanced workflows for experts.
Future research should explore how vibe coding practices can be integrated into training for non-software developers, so that educational interventions complement tool design and enhance baseline QA competence.

\section{Threats to Validity}
\label{ThrToVal}



\noindent \textbf{Internal Validity:} To address possible \textit{search bias} in our GLR, we first designed a comprehensive search strategy. We employed a detailed search strategy with iterative pilot searches to refine the search terms (see Section \ref{SearchStrDev}). 
This approach broadened coverage and reduced search string bias, but given the diffuse and evolving nature of grey literature, we cannot claim to have captured all relevant sources.
We also defined a set of \textit{inclusion and exclusion criteria} (see Section \ref{IncExc}) and applied them consistently during the data filtration to minimize any \textit{selection bias}.

To minimize \textit{quality bias}, we applied Garousi et al.’s five-dimension quality checklist \cite{garousi2019guidelines} (see Section \ref{QAasst}), excluding sources scoring below 10/15. This reduces the likelihood that anecdotal or promotional content disproportionately influenced the findings, although there is still some risk of quality bias.


With regards to \textit{data extraction bias}, to avoid emphasizing certain quotes or themes, we used a standardized extraction template and applied Braun and Clarke \cite{braun2006using} structured thematic analysis steps (see Section \ref{DataAnalysis}). Coding was conducted primarily by the first author, with discussions and consensus building among the other authors. Pilot extractions and iterative refinement helped calibrate interpretation in findings. However, thematic coding inevitably contains subjective judgment, so our findings should be understood as patterns drawn from the data rather than precise measurements.


\noindent \textbf{External Validity:} 
Our study is based on \NumberIncluded{} grey literature sources, which reflect the experiences of practitioners who chose to share their stories online. This introduces \textit{self-selection bias}, those represented may not reflect all vibe coders. We mitigated this by ensuring diversity in high-quality source types (blogs, forums, and media articles) and user roles (novices and experienced users) from various industries. However, the findings cannot be considered generalizable to all vibe coders. Instead, they should be interpreted as indicative of common patterns in reported experiences, not as representative of the entire population of vibe coders.

\noindent\textbf{Construct Validity:}
To establish a consistent data extraction basis for our GLR, we utilized a standard data extraction template to ensure uniformity. We refined our data extraction methods after several pilot tests. Reviews and discussions with other co-authors were carried out for the extracted data process, resolving discrepancies through consensus. Behavioral unit extraction was conducted mainly by the first author. To increase reliability, a subset of units was cross-validated by another co-author, minimizing the likelihood of inaccurate conclusions. This reduced, but did not eliminate, the risk of subjective bias and inaccurate conclusions.

\section{Conclusion}
\label{Conclusion}

This study provides the first empirical investigation of how users actually engage in vibe coding, especially outside formal development settings. By systematically analyzing \NumberIncluded{} grey literature sources containing \NumBehavioralUnits{} firsthand behavioral units, we uncovered why users engage in vibe coding, what they experience while doing so, how they perceive the quality of \aigc{}, and what practices they apply to review or test the code.

Our findings reveal a speed–quality trade-off in vibe coding; vibe coders, particularly those without software development experience, are enabled to create usable applications quickly, which often comes at the expense of verification and maintainability limitations. The widespread use of delegated QA to AI, uncritical trust, and skipped testing highlights how vibe coding outputs are vulnerable. Moreover, the emergence of \textit{vulnerable developers}(who are capable of building but unable to debug) highlights the risks of democratizing software creation without corresponding investments in quality practices. For tool designers, our results suggest including QA feedback, visualizing code quality indicators, and providing inline explanations to mitigate the impact of uncritical trust. 
For teams adopting AI-assisted development, cautious use of vibe coding in production systems is recommended. Adhering to organizations' guardrails that require review processes and maintaining debugging skills.

\AItools{} support will play a bigger role in software development in the future. How we respond to the behavioral patterns identified in this study will determine whether the future improves or degrades software quality. The \AItools{} exist; the challenge lies in using them wisely.


\section*{Data Availability Statement}
We provide our data and detailed analysis in 
 \cite{fawzy_2025_17188020}

\bibliographystyle{ACM-Reference-Format}
\bibliography{main.bib}

\end{document}